\documentclass[twocolumn,showpacs,preprintnumbers,nofootinbib,prd,superscriptaddress,10pt,floatfix]{revtex4-2}

\usepackage{graphicx}
\usepackage{bm}
\usepackage{orcidlink}
\usepackage{gensymb}
\usepackage{natbib}
\usepackage{booktabs}
\usepackage{amsmath}
\usepackage[caption=false]{subfig}
\begin{document}

\preprint{APS/123-QED}

\title[Peep Signal Confusion Noise]{Gravitational Wave Peep Contributions to Background Signal Confusion Noise for LISA}

\author{Daniel J Oliver \orcidlink{0000-0002-7374-6925}}
\thanks{Email: oliverda@oregonstate.edu}%
\affiliation{Oregon State University, 1500 SW Jefferson Ave, Corvallis, OR 97331, USA}
\affiliation{Department of Physics, University of Arkansas, Fayetteville, AR 72701, USA}
\affiliation{Arkansas Center for Space and Planetary Sciences, University of Arkansas, Fayetteville, AR 72701, USA}%

\author{Aaron D Johnson \orcidlink{0000-0002-7445-8423}}
\affiliation{Theoretical Astrophysics Group, California Institute of Technology, Pasadena, CA 91125, USA}

\author{Lena Janssen \orcidlink{0009-0004-3049-9436}}
\affiliation{%
 Department of Physics and Astronomy, University of Nebraska-Kearney, Kearney, NE 68849, USA}%
 
\author{Joel Berrier}
\affiliation{%
 Department of Physics and Astronomy, University of Nebraska-Kearney, Kearney, NE 68849, USA}%
 
\author{Kostas Glampedakis \orcidlink{0000-0003-1860-0373}}
\affiliation{Departamento de Física, Universidad de Murcia, Murcia E-30100, Spain}

\author{Daniel Kennefick \orcidlink{0000-0002-5219-456X}}
\affiliation{%
 Department of Physics, University of Arkansas, Fayetteville, AR 72701, USA}
 \affiliation{Arkansas Center for Space and Planetary Sciences, University of Arkansas, Fayetteville, AR 72701, USA}%

\begin{abstract}
Two-body gravitational interactions will occasionally lead to a stellar-mass compact object entering a very highly eccentric orbit around a massive black hole at the center of a galaxy. Gravitational radiation damping will subsequently result in an extreme mass ratio inspiral. Much of the inspiral time of these events is spent with the compact object on a long-period orbit, with a brief burst of gravitational wave emission at periapsis firmly in the mHz band. Burst orbits have been previously modeled as parabolic, with a focus on extreme examples that could be detectable by space-based gravitational wave detectors. This work focuses on the recurring bursts called ``peeps". Peeps are not likely to be individually resolvable; however, it is also important to consider them as possible sources of signal confusion noise because they do generate a signal within the LISA band with every pericenter passage. To account for peeps, we must utilize estimates for EMRI capture parameters along with tracking the massive black hole population out to a redshift of 3 using the Illustris Project. Then, this population is combined with an EMRI formation rate to estimate the number of EMRI events per unit volume for LISA. In this study, we model four different assumptions for the gravitational wave background produced by these highly eccentric peeps. We find that with our two most likely backgrounds, the signal may result in a slight rise of the LISA noise floor (SNR $\sim 0.3-2.4$); however, in two more abundant cases, the background generated by these sources would be detectable on their own and likely obscure many potentially detectable sources (SNR $\sim77-145$).
\end{abstract}

\maketitle

\section{Introduction\protect}

Capture events of stellar-mass compact objects (COs) by massive black holes (MBHs) at the center of galaxies will be an important gravitational wave source for the planned ESA-NASA space-based gravitational wave detector LISA, the Laser Interferometer Space Antenna \cite{amaro-seoane_laser_2017, barack_confusion_2004, glampedakis_zoom_2002, babak_kludge_2007, babak_science_2017}. These capture events of COs orbiting around MBHs are called Extreme Mass Ratio Inspirals (EMRIs) and are formed from multi-body gravitational interactions that result in one of the COs entering a very highly eccentric orbit around the MBH evolving and inspiraling over very long periods due to the emission of gravitational waves in the millihertz bandwidth \cite{amaro-seoane_laser_2017, babak_kludge_2007, babak_science_2017, barack_confusion_2004, barack_lisa_2004, bonetti_gravitational_2020,glampedakis_approximating_2002, Oliver_2024, gairTestingGeneralRelativity2013a,ryan1995,barack2007,gair2010, gair_prospects_2017}. 

EMRI orbits typically have mass ratios between $q \sim 10^{-7}-10^{-4}$, where the mass ratio is defined as $q\equiv\mu/M$, with $\mu$ as the CO mass and $M$ is the MBH mass. EMRIs spend most of their ``life" in long-period, highly eccentric orbits slowly inspiraling for potentially millions of years \cite{Oliver_2024, babak_science_2017, seoane2024mono}. 


Many previous studies have looked at the possibility of a gravitational wave background in LISA \cite{barack_confusion_2004,romanoDetectionMethodsStochastic2017a, Caprini_2019, bonetti_gravitational_2020, karnesis2021, Flauger_2021, pozzoli2023computation}. Recent studies, such as those of Pozzoli et al. (2023), have examined EMRIs up to an eccentricity of $0.9$ (with estimates for SNRs above $e=0.9$), many of which are potentially detectable by LISA on their own. However, this does not cover the entire spectrum of EMRIs. The very highly eccentric (near-parabolic) sources, where EMRIs spend much of their early inspiral time, must also be considered, as these sources may add up to a potentially unresolvable gravitational wave background that may obscure otherwise detectable sources for LISA \cite{Oliver_2024}. One recent study has even argued that there could be hundreds to thousands of these sources in a Milky Way-type galaxy \cite{seoane2024mono}.

At capture, these orbits are highly eccentric, long-period, and much too low frequency for the entire waveform to be detectable in the LISA band. However, there is a burst of emission of gravitational radiation at periapsis, which is firmly in the LISA bandwidth \cite{Oliver_2024}. These events are called Extreme Mass Ratio Bursts (EMRBs). They are defined by a pericenter passage with a timescale less than $10^5$ seconds and have been traditionally modeled as parabolic orbits where $e = 1$ \cite{rubboEventRateExtreme2006, hopmanGravitationalWaveBursts2007,yunesRelativisticEffectsExtreme2008,toonenGravitationalWaveBackground2009, berryObservingGalaxyMassive2013a, berryExtrememassratioburstsExtragalacticSources2013a, berryExpectationsExtrememassratioBursts2013a,   fanExtrememassratioBurstDetection2022}. 

If the eccentricity is just below unity, then the CO will continue on its orbit beyond the initial burst of emission, losing angular momentum, and beginning an inspiral towards the MBH. It is these signals that are the focus of this paper. We are using the word gravitational wave ``peep" to refer to the short, high-pitched signal amidst a low-frequency background. A peep differs from a burst in that it will recur, typically many times, at intervals of one orbital period of the source \cite{Oliver_2024}.

These highly eccentric orbits have other unique qualities, such as exhibiting Zoom-Whirl behavior, which is described by a long-period gravitational wave feature with a much shorter periodic feature at periapsis \cite{glampedakis_approximating_2002}. These orbits are also known as homoclinic orbits \cite{levinGravityWavesHomoclinic2000}. These ``peep" signals produce repeated bursts (or partial whirls) at periapsis throughout most of their orbit, evolving in chirp-like fashion both in frequency and amplitude. Most of the amplitude of the inspiral is the result of these peeps, as the zoom portion of the orbit results in a very low amount of gravitational wave emission. One interesting aspect though, is that the amplitude evolution of the peep is very insignificant until the very end of the inspiral near merger, this is due to the periapsis distance remaining roughly constant, while the apoapsis is greatly decreasing with each passage, and it isn't until the orbit is nearly circular that we see significant amplitude evolution due to the change in periapsis \cite{Oliver_2024}.

The structure of this paper is as follows. Section \ref{sec:methods} will discuss the waveform model, rates, SMBH mass function, and the parameter space for peeps. Section \ref{sec:results} will discuss the four backgrounds that we analyze and their creation. Section \ref{sec:conc} is where we draw our final conclusions. Throughout this paper we use geometric units where $G=c=1$.

\section{\label{sec:methods}Methods}
\subsection{Numerical Kludge Model}

These highly eccentric (near-parabolic) orbits are computationally expensive and extremely difficult to model in a generic way, with a spinning MBH and eccentric, inclined orbits. The Numerical Kludge model was chosen for this study as it captures many of the relativistic effects that appear in ``peep" orbits. The specific code utilized in this study is the time-domain Fortran code developed and described in Babak et al. (2007) \cite{babak_kludge_2007}, with a modification to include $h_\times$ polarization.

Since these orbits also do not evolve appreciably within LISA's 4-year observing window, we do not model the orbits with orbital evolution. This model begins by producing an inspiral trajectory in phase space defined by the orbital energy ($E$), axial angular momentum ($L_z$), and the Carter constant ($Q$). The code then numerically integrates the Kerr geodesic equations to obtain the Boyer-Lindquist coordinates $\{t,r,\theta,\phi\}$ of the compact object, which are then projected onto Cartesian coordinates and used in the quadrupole formula to construct the gravitational waveform\footnote{For a more detailed description of the Numerical Kludge model, see Babak et al. (2007) \cite{babak_kludge_2007}}. The NK model is a semi-relativistic model which has overlaps with Teukolsky waveforms of 95\% or higher in the regions of parameter space we are interested in ($r_p \geq 5M$ corresponding to $p \gtrsim 10M$ for peep eccentricities) \cite{babak_kludge_2007, gair_semi-relativistic_2006, ruffini_semi-relativistic_1981}.

\subsection{Rates}

There are three ways to approach the rates for these highly eccentric EMRIs. The first of which is to consider previous studies of EMRBs, which for a Milky Way type galaxy using eLISA, it was estimated that there will be on the order of $1\ \mathrm{yr}^{-1}$ detectable, and outside of our galaxy will be an order of magnitude lower \cite{rubboEventRateExtreme2006, hopmanGravitationalWaveBursts2007, berryExpectationsExtrememassratioBursts2013a, berryExtrememassratioburstsExtragalacticSources2013a, berryObservingGalaxyMassive2013a}. For TianQin out to a distance of $100 \mathrm{Mpc}$, there would be $\sim 7\ \mathrm{yr}^{-1}$ \cite{fanExtrememassratioBurstDetection2022}, these rates are described in Table~\ref{tab:EMRBRate}.
Another approach is to consider rates based on the formation of EMRI signals that are likely to eventually merge, one example is described in Babak et al. (2017)\cite{babak_science_2017}. \par 
A final alternative to rates is to look at large numbers of early-stage EMRIs, which are likely to re-scatter. A recent study looked at these highly eccentric EMRIs in the galactic center and the detectability rates of a galactic foreground using Keplerian orbits and found that there could be hundreds to thousands of these signals in the galactic center \cite{seoane2024mono}.

\begin{table}
  \caption{\label{tab:EMRBRate} Detectable EMRB/Early EMRI Event Rates for the Milky Way galaxy and beyond.}
  \begin{tabular}{@{}p{\dimexpr\columnwidth-2\tabcolsep}@{}}
    \toprule
    \textbf{Detectable EMRB Rates} \\
    \midrule
    Milky Way\footnotemark[1]: $\sim 15\ \mathrm{yr}^{-1}$\\
    Virgo Cluster\footnotemark[1]: $\sim 3\ \mathrm{yr}^{-1}$\\
    Milky Way\footnotemark[2]: $\sim 1.1\ \mathrm{yr}^{-1}$\\
    Virgo Cluster\footnotemark[2]: $\sim 0\ \mathrm{yr}^{-1}$\\
    M32, NGC 4945, NGC 4395\footnotemark[3]: $\sim 0.2\ \mathrm{yr}^{-1}$\\
    Milky Way\footnotemark[4]: $\sim 1\ \mathrm{yr}^{-1}$\\
    Up to 100 Mpc away\footnotemark[5]: $\sim 7\ \mathrm{yr}^{-1}$\\
    Milky Way\footnotemark[6]: $\sim 100\mathrm{s}$ to $1000\mathrm{s}\ \mathrm{yr}^{-1}$\\
    \bottomrule
  \end{tabular}
  \footnotetext[1]{Rubbo et al. (2006) (SNR $\geq 5$) \cite{rubboEventRateExtreme2006}}
  \footnotetext[2]{Hopman et al. (2007) (SNR $\geq 5$) \cite{hopmanGravitationalWaveBursts2007}}
  \footnotetext[3]{Berry \& Gair (2013b) (SNR $\geq 10$) \cite{berryExtrememassratioburstsExtragalacticSources2013a}}
  \footnotetext[4]{Berry \& Gair (2013c) (SNR $\geq 10$) \cite{berryExpectationsExtrememassratioBursts2013a}}
  \footnotetext[5]{Fan et al. (2022) (SNR $\geq 10$) \cite{fanExtrememassratioBurstDetection2022}}
  \footnotetext[6]{Amaro-Seoane et al. (2024) (SNR $\geq 10$) \cite{seoane2024mono}}
\end{table}

\begin{figure}
    \centering
    \includegraphics[width=\columnwidth]{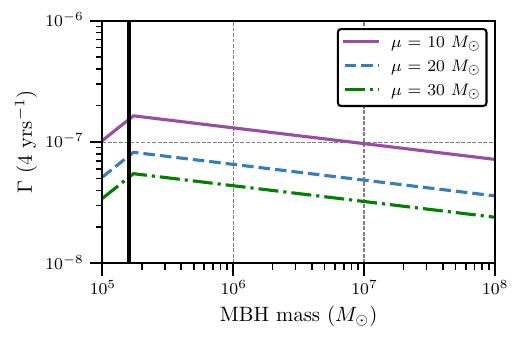}
    \caption{Calculated following Babak et al. (2017) \cite{babak_science_2017} using Eqs. 23-31, with a fixed plunge rate, a variable CO mass, and adjusted to be a rate $\Gamma$ given a $4$-year window for a given MBH mass. This rate is assuming that there are 10 plunges per formed EMRI, thus renormalizing the EMRI rate to account for sources that directly plunge after capture. The MBH mass limits were set based on our population model described in \S\ref{sec:massfunc}, with the vertical black bar denoting the lower bound on MBH mass. We also include three separate rates based on different CO masses. In our study, we consider a uniform distribution of CO masses between these values (See Table~\ref{tab:Param_dist}).}
    \label{fig:babakfig4}
\end{figure}

The rate used in our study is based primarily on the work described in Babak et al. (2017) \cite{babak_science_2017}\footnote{For a full discussion of how the rates are calculated, please see Babak et al. (2017) \cite{babak_science_2017}}. In our study, we take the median result from the rates described in Equations 23-31, assuming a ratio of 10 plunges per formed EMRI ($N_p=10$). Figure~\ref{fig:babakfig4} was generated following these methods and shows the EMRI formation rates adjusted to show the rate over a 4-year window for a given MBH mass based on our population model and for three separated CO masses. The results of method this align with those presented in Figure 4 of Babak et al. (2017) \cite{babak_science_2017}. 

\subsection{Signal Confusion Noise}

Several studies have been conducted on EMRI signal confusion noise \cite{barack_confusion_2004, bonetti_gravitational_2020, racine_gaussianity_2007, chua_non-local_2022, pozzoli2023computation} as well as EMRB signal confusion noise \cite{toonenGravitationalWaveBackground2009, fanExtrememassratioBurstDetection2022} on various detectors such as eLISA, LISA, and TianQin. However, very little work has been done on the very highly eccentric EMRIs between the two extremes of the parabolic EMRBs and the low eccentricity EMRIs. These signals must be considered, as they exist primarily in the lower frequency range of the LISA bandwidth \cite{Oliver_2024}. Pozzoli et al. (2023) \cite{pozzoli2023computation} has provided one of the most comprehensive studies thus far on a stochastic gravitational wave background from EMRIs. The results from this work showed that there is significant SNR from a GWB after removing the individually resolvable sources. In this remaining GWB lies the highly eccentric signals ($e>0.9$) which were removed from their study due to waveform model limitations. The in depth examination of these signals is the topic of this paper. 

These repeated bursts, or ``peep" signals, as seen in Table~\ref{tab:EMRBRate}, are not likely to be detectable on their own, with $\sim 1\mathrm{yr}^{-1}$ EMRB being detectable, there are numerous signals that remain undetectable. If we then take into account the peep signals, which have larger amplitudes in the frequency domain due to repeated bursts, then there are far more sources to consider and thus more likely undetectable sources that will add together, contributing to an overall background signal confusion noise \cite{Oliver_2024}. To construct this background, we will first need to determine a SMBH mass function that we can combine with our formation rates from the previous section to generate our peep background.

\subsection{\label{sec:massfunc}SMBH Mass Function}

To create our SMBH mass function and to be able to quantify the evolution of that function over a wide range of cosmic history, we have elected to extract our mass function from a cosmological simulation. We chose the Illustris-1 simulation \cite{vogelsberger_2014a, vogelsberger_2014b, sijacki_illustris_2015} as a well-studied, publicly available, large-scale and high-resolution hydrodynamic cosmological simulation, with a wide range of astrophysical processes, e.g., gas cooling, AGN feedback from growing MBHs and exploding supernovae, and radiation proximity effects for AGN. Illustris also includes massive black hole mass evolution over a wide range of cosmic time. Illustris is a hydrodynamic simulation of galaxy formation in a co-moving cube with side length $75\ \mathrm{ Mpc }\ h^{-1}$. Illustris uses the WMAP-9 cosmology \cite{Hinshaw2013}, with $\Omega_M = 0.2726$, $\Omega_{\Lambda} = 0.7274$, $\Omega_b = 0.0456$, $\sigma_8 = 0.809$, and $h = 0.704$. We have adopted the cosmology used in the Illustris simulations for necessary calculations.

In this simulation, massive black holes are created when a dark matter halo gains a mass that exceeds $7.1 \times 10^{10} M_{\odot}$ if that halo does not already have a massive black hole due to another event. These initial massive black holes have a starting mass of $10^{5.0}\ h^{-1} M_{\odot}$. These seed massive black holes may continue to grow through mergers with other massive black holes, or through gas accretion. The gas accretion is based on an Eddington-limited Bondi–Hoyle–Lyttleton model \cite{DiMatteo2005, Springel2005}. Due to the starting massive black hole mass, we are thus limited to a lower mass of $1.42 \times 10^{5} M_{\odot}$. Constraints are also placed on the maximum massive black hole mass due to the size of the simulation, with further constraints of our own due to mass ratio for EMRIs. Combined this provides a reasonable measurement of the MBH population between $\log_{10}(M) = 5.2 - 8$. 

Massive black hole data can be obtained from Illustris in two different ways. Individual simulation snapshots contain massive black hole data, but there is also a black hole details file that includes additional information on the massive black hole population between the available simulation snapshots. This allows for additional fine-scale information on the massive black hole population with some exceptions. The black hole details files are corrupted and information is missing between $z = 0.14$ and $z = 0.38$. Over this range, only data from the individual simulation snapshots is used. 

\begin{figure}[t]
  \centering
  \subfloat[]{\includegraphics[width=\columnwidth]{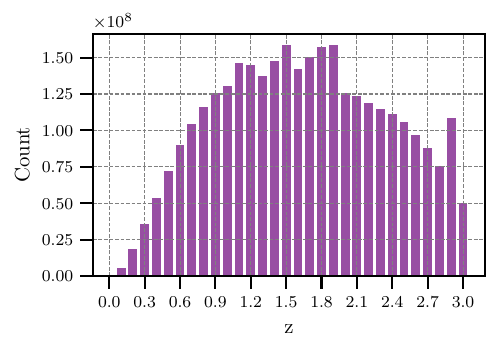}}\\[-0.5\baselineskip]
  \subfloat[]{\includegraphics[width=\columnwidth]{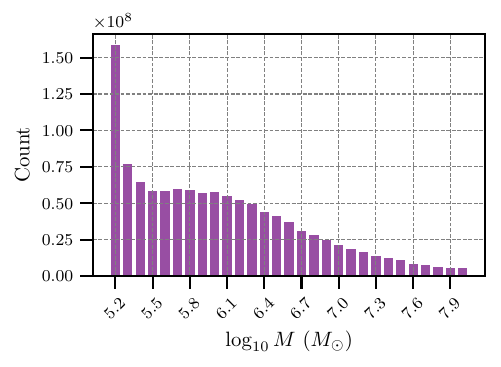}}
  \caption{(top): 2D histogram of mass function showing the count of MBHs versus redshift. (bottom): 2D histogram of mass function showing the count of MBHs versus $\log_{10}M$ in solar mass. }
  \label{fig:redshift_logmass}
\end{figure}


The MBH mass functions were calculated in logarithmic bins of $0.1~ \mathrm{dex}$. The mean mass of the massive black holes in the bin was calculated along with the number density, $n$, of black holes in that mass bin per unit of comoving volume. In order to estimate the final number of massive black holes at a given $\log_{10}(M)$, we linearly interpolate $\log_{10}(n)$ with respect to the $\log_{10}(M)$. In cases where the number of massive black holes between a step in the interpolation would be zero, we used a $\log_{10}(M)$ vs. $n$ instead. This interpolation was performed at each of the snapshots between a redshift $z=0$ and $z=3.01$, represented by snapshots $60$ through $135$ in the Illustris data. This data set is augmented with the additional results from the black hole details file. The mass function is calculated at all available times from the simulation data. Finally, a linear interpolation is used to calculate the number density of massive black holes at a given mass at higher redshifts. In this case, the $\log_{10}(n)$ is calculated in each bin by interpolating between the scale factors, $a$, of the available data.

Finally, the number of expected black holes at each given mass bin was calculated using the interpolated mass functions and numerical integration over the comoving volume. First, the number of black holes in a local volume of the universe was calculated. The $z = 0$ mass function provides the number density of massive black holes in a spherical shell that extends to a physical radius corresponding to $z=0.05$ in our selected cosmology. After that, concentric spherical shells were integrated over, with redshift increments of $dz=0.1$. In each shell, the mass function for the mean redshift of the shell was used to estimate the number of massive black holes in that volume. In this case, we are assuming the Illustris mass functions are a representative sample over each of these redshift ranges of the number density of SMBHs. In each redshift range, we can estimate the number of expected massive black holes in each mass and redshift bin. Figure~\ref{fig:redshift_logmass}
shows the 2D representations of the calculated MBH mass function, showing the count of MBHs versus redshift and $\log_{10}M$. Due to the expected mass ratios of EMRIs, we have limited our MBH mass range to $\log_{10}(M) = 5.2 - 8.0$ \cite{vogelsberger_2014a,vogelsberger_2014b,sijacki_illustris_2015}.

\subsection{Parameters}

EMRIs are fully defined in a 16-dimensional parameter space:\\ \{$a, a_2, p_0, e_0, \iota_0, M, \mu, \theta_s, \phi_s, \theta_k, \phi_k, \Phi_{\phi,0}, \Phi_{\theta,0}, \Phi_{r,0}, \chi, d_l$\} \cite{FEW1,FEW2}. For this study and our use of the numerical kludge model, we can reduce the necessary parameters for our waveform model down to a 12-dimensional space: \{$a, p_0, e_0, \iota_0, M, \mu, \theta_s, \phi_s, \theta_k, \phi_k, \chi, d_l$\}. The NK model uses the observation angles $\theta_s$ and $\phi_s$ for the orbit's orientation with respect to an observer. The sky location coordinates $\theta_k, \phi_k$ are introduced in the LISA response function along with the polarization angle $\chi$. We assume that the CO is on a prograde orbit, so $a > 0$, and that the CO is non-spinning, thus $a_2=0$. We also assume that the initial phase of the fundamental frequencies is fixed, so we can safely ignore \{$\Phi_{\phi,0}, \Phi_{\theta,0}, \Phi_{r,0}$\}. We also use redshift $z$ as our input and convert to the luminosity distance $d_l$ utilizing \texttt{Astropy}\cite{astropy:2013,astropy:2018,astropy:2022} for our chosen cosmology for this conversion. The final parameters along with their distributions for the four backgrounds described in \S\ref{sec:results} are shown in Table~\ref{tab:Param_dist} and are based in part by work done in Oliver et al. (2024) \cite{Oliver_2024}. 

\begin{table*}[ht!]
    \centering
    \caption{Gravitational Wave Peep parameters and distributions for the four background assumptions used in this study. Based on Oliver et al. (2024) \cite{Oliver_2024}.}
    \begin{tabular}{l l l l}
        \toprule
        \textbf{Parameter} & \textbf{Description} & \textbf{Distribution} \\
        \midrule
        $a/M$ & Dimensionless spin of the MBH & $\mathcal{U}[0.8, 0.998]$  \\
                $(p_0/M)_{1}$ & Initial semi-latus rectum (Background 1) & $\mathcal{U}[15, 120]$  \\
        $(e_0)_{1}$ & Initial orbital eccentricity (Background 1) & $1 - \mathcal{U}[10^{-5}, 10^{-2}]$  \\
        $(p_0/M)_{2,3,4}$ & Initial semi-latus rectum (Background 2,3,4) & $\mathcal{U}[8, 120]$  \\
        $(e_0)_{2,3,4}$ & Initial orbital eccentricity (Background 2,3,4)& $1 - \mathcal{U}[10^{-5}, 10^{-1}]$  \\
        $\iota_0$ & Initial orbital inclination & $\mathcal{U}[0^\circ, 180^\circ]$  \\
        $M$ & MBH mass in $M_\odot$ & $[10^{5.2}, 10^8]$  \\
        $\mu$ & CO mass in $M_\odot$ & $\mathcal{U}[10, 30]$ \\
        $\theta_s$ & Polar sky location angle (degrees) & $\mathcal{U}[0^\circ, 180^\circ]$  \\
        $\phi_s$ & Azimuthal sky location angle (degrees) & $\mathcal{U}[0^\circ, 180^\circ]$  \\
        $\theta_k$ & Polar angle describing spin of MBH (degrees) & $\mathcal{U}[0^\circ, 180^\circ]$  \\
        $\phi_k$ & Azimuthal angle describing spin of MBH (degrees) & $\mathcal{U}[0^\circ, 180^\circ]$  \\
        $\chi$ & Polarization Angle (degrees) & $\mathcal{U}[0, 180]$  \\
        $z$ & Redshift to source & $[0, 3]$  \\
        \bottomrule
    \end{tabular}
    \label{tab:Param_dist}
\end{table*}

\subsection{LISA Response Function}

LISA employs time-delay interferometry (TDI), which is a form of synthetic interferometry that intelligently cancels out otherwise overwhelming laser noise by implementing linear combinations of delayed phase measurements of the input data. Each of the three satellites in the LISA constellation reads in data and transmits signals between the other two satellites, recording the phase of the interference between its local laser and the received signal, which contains a time-delayed copy of the distance spacecraft's laser noise plus a gravitational wave signal. Through these linear combinations, we receive outputs in three data channels, X, Y, and Z, which are combined in post-processing into orthogonal channels A, E, and T in such a way that the signal present in T is significantly diminished, resulting in a so-called null channel \cite{vallisneri_2005, Vallisneri_2021, cutler1998, mcnamara2000, lisaongpu1, colpi2024lisa}. Due to T being the null channel, we will only include the A and E channels in our resulting plots. 

TDI can be generalized to accommodate the constellation dynamics, i.e., the motion of the individual satellites that make up LISA, such as their changing rotation and arm lengths \cite{colpi2024lisa}. In this study, we utilized the \texttt{fastlisaresponse} \cite{lisaongpu1,lisaongpu2} code, with 2nd generation TDI and the ESA trailing orbits option, which uses numerically generated orbits of the three satellites provided by ESA \cite{Martens_2021}. The ESA trailing orbits account for all relevant bodies in the solar system, such as the planets and moons, which may impact the orbit. The orbits are also optimized to try to minimize the ``breathing" that occurs in the constellation. These orbits are computed in \texttt{fastlisaresponse} with \texttt{LISAOrbits} \cite{bayle_2022_6412992}.\par

\section{Results \label{sec:results}}

For the analysis of the gravitational wave background, we use the \texttt{LISAAnalysisTools} \cite{michael_katz_2024_10930980} suite of codes to generate a sensitivity curve for LISA. For the sensitivity curve, we used the A1TDISens and E1TDISens for the A and E channels, respectively, assuming 4 years of stochastic confusion noise from white dwarf binaries with the \texttt{scirdv1} LISA model. \par


Using the rates in Figure~\ref{fig:babakfig4}, the SMBH mass function described in Section~\ref{sec:massfunc}, the parameters and distributions in Table~\ref{tab:Param_dist}, and the LISA sensitivity curve/analysis tools from \cite{michael_katz_2024_10930980} we can create our catalog of peep waveforms. For comparison, we have created four different peep backgrounds based on different assumptions: 
\begin{enumerate}
    \item Rates from Figure~\ref{fig:babakfig4} with current estimates for capture parameters along with an assumption of 1 or fewer highly eccentric EMRIs per MBH during LISA's observing window are captured and begin their inspiral with at least one pericenter passage within 4-years resulting in a burst of gravitational waves.
    \item Rates from Figure~\ref{fig:babakfig4} with adjusted estimates for capture parameters to better connect with previous gravitational wave background calculations from literature \cite{pozzoli2023computation}. Adjusting capture semi-latus rectum to $8M \leq p_0 \leq 120M$ and eccentricity to $0.9 \leq e_0 \leq 0.999999$ still with the assumption of 1 or fewer highly eccentric EMRIs per MBH during LISA's observing window.
    \item Rates from Figure~\ref{fig:babakfig4} with adjusted estimates for capture parameters to better connect with previous studies. Adjusting capture semi-latus rectum to $8M \leq p_0 \leq 120M$ and eccentricity to $0.9 \leq e_0 \leq 0.999999$ with the added assumption that for each MBH which observed highly eccentric EMRIs during LISA's observing window, there were 1000 of that same signal. This allows us to consider the more abundant case presented in \cite{seoane2024mono}.
    \item Rates from Figure~\ref{fig:babakfig4} with adjusted estimates for capture parameters to better connect with previous studies. Adjusting capture semi-latus rectum to $8M \leq p_0 \leq 120M$ and eccentricity to $0.9 \leq e_0 \leq 0.999999$. This background estimate attempts to account for those signals that formed before the 4-year window of the other three backgrounds. To approximate this contribution, we use the orbital periods of each peep in our catalog. The maximum orbital period within our parameter distribution is $\sim10^5$ years, which we take as the reference timescale $\mathcal{T_{\mathrm{ref}}}$. We then look only at peeps in our catalog with orbital periods ($\mathcal{P}_{\mathrm{orb}}$) $\geq 4$ years. Systems with shorter periods are excluded, as they are likely to evolve into individually resolvable signals within $\sim10^5$ years. We then obtain an estimated multiplicative factor ($\mathcal{F}$) for the total number of additional orbits to consider for our background of peeps. This is done using Equation~\ref{eqn:factor}.  This procedure yields an estimated factor of 3545 additional orbits per source using data from \textbf{Assumption 2}. To account for interference among signals, we scale the background contribution from \textbf{Assumption 2} by $\sqrt{3545}$, providing a reasonable estimate for the upper limit on highly eccentric EMRIs.
    \begin{equation} \label{eqn:factor}
        \mathcal{F} = \frac{\mathcal{T}_{\mathrm{ref}}}{\mathcal{P}_{\mathrm{orb\geq4 \mathrm{yrs}}}}
    \end{equation}
\end{enumerate}

We also make some general assumptions which are included in each of the four backgrounds: assuming fixed distances to the EMRI systems based on the redshift and our chosen cosmology, a fixed plunge to EMRI formation rate set to $N_p = 10$ (comparable to M1-M6 and M9-M10 from Babak et al. (2017) \cite{babak_science_2017}, as well as assuming a median amount of MBHs at each redshift and given mass value. This means that we are omitting the uncertainty in the MBH counts.

\begin{figure}[hb!]
    \centering
    \includegraphics[width=\columnwidth]{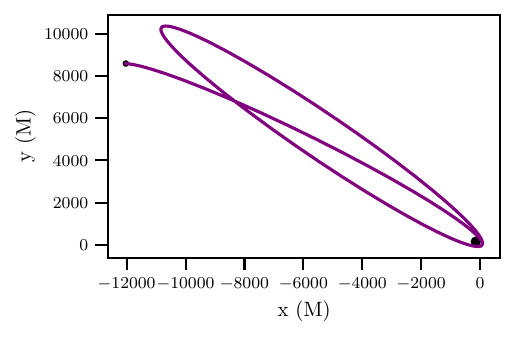}
  \caption{Orbit of the first 4 months of an orbit of a peep with the same parameters as Figure~\ref{fig:test_wave_char}. The orbit is shown in Cartesian coordinates with units of $M$. In this figure, the small body begins its orbit at periapsis. Here one can see the orbital precession due to the burst, which occurs after the small body completes its first orbit and passes periapsis again. This repeats with each pericenter passage forming the peep signal \cite{Oliver_2024}.}
  \label{fig:peep_cartesian}
\end{figure}

\begin{figure*}[t]
  \centering
  \subfloat[]{\includegraphics[width=0.95\textwidth]{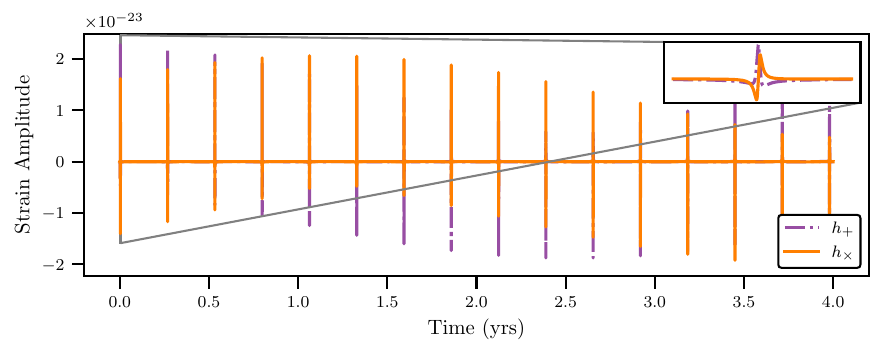}}\\[-0.75\baselineskip]
  \subfloat[]{\includegraphics[width=0.95\textwidth]{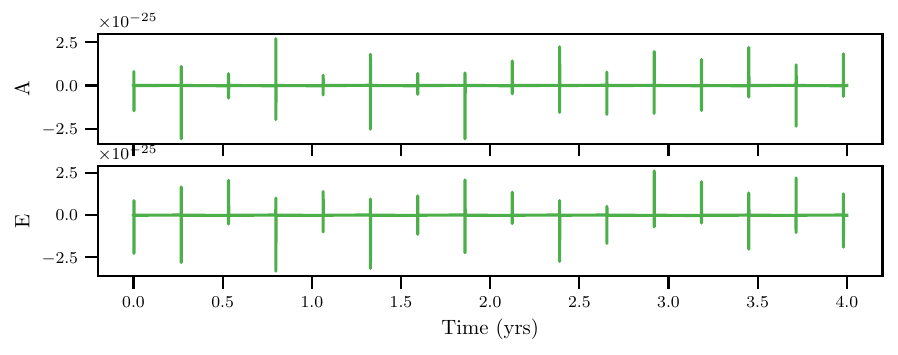}}\\[-0.75\baselineskip]
  \subfloat[]{\includegraphics[width=0.95\textwidth]{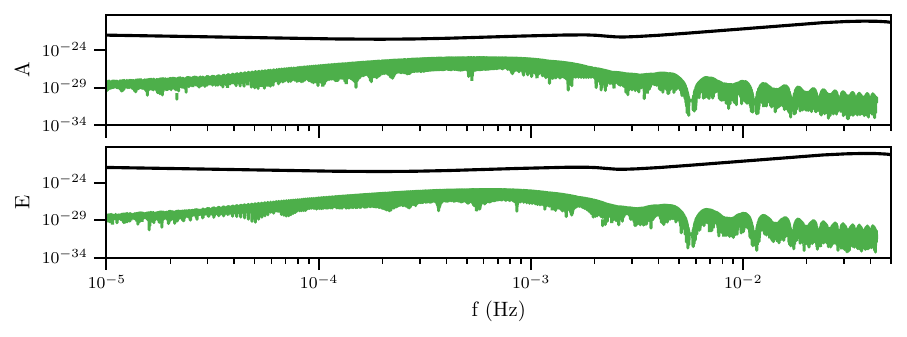}}
  \caption{
  (top): Peep waveform with parameters $a=0.815535M$, $p_0=113.194144M$, $e_0=0.992424$, $\iota_0=2.363436^\circ$, $log_{10}\ M=5.2\ M_\odot$, $\mu=18.152765\ M_\odot$, $\theta_k=0.581082^\circ$, $\phi_k=14.558132^\circ$, $z=0.6$, and $dt=15 \mathrm{s}$ at a sky location of $\theta_s=0.980574^\circ$, $\phi_s=5.229798^\circ$ generated using the above methodology. The waveform depicts 4 years of a highly eccentric EMRI in the detector frame. On the y-axis is the strain amplitude of the gravitational wave signal. In purple dashed lines is the $h_+$ polarization, and in orange solid lines is the $h_\times$ polarization. The inset figure is zoomed in on the first burst in the repeated burst (peep) signal. (middle): The A and E channel outputs after passing the above waveform through \texttt{fastlisaresponse}. {bottom}: Characteristic strain for A and E channel plotted over the A1TDISens and E1TDISens sensitivity curves, respectively. The SNR for the A and E channels are $\sim 0.0012$ and the combined SNR is $\sim 0.0017$, which is undetectable on its own.
  }
  \label{fig:test_wave_char}
\end{figure*}


\subsection{\label{sec:peepsamples} Peep Sampling}





We begin by cycling through the data shown in Figure~\ref{fig:redshift_logmass} for each redshift and MBH mass. We then take a random draw of a CO mass $\mu$ using the distribution in Table~\ref{tab:Param_dist} and determine the probability of an EMRI occurring during LISA's observing window. We do this by implementing a probability mass function \texttt{scipy.binom.pmf}\cite{scipy}. Since we are only concerned with whether at least one EMRI occurs, the probability simplifies to:
\begin{equation} \label{eqn:prob_occ}
    P(\mathrm{occurrence}) = 1 - binom.pmf(0, 1, \Gamma_{4yr}) 
\end{equation}
where $\Gamma_{4yr}$ is the expected number of events in a 4-year window from Figure~\ref{fig:babakfig4}. We compute this probability for each MBH to determine if at least one EMRI occurs within LISA's observation time. We then generate a uniformly distributed random variable $X$ (using \texttt{numpy.random.rand()} \cite{numpy}) and compare it to $P(\mathrm{occurrence})$. If $X \leq P(\mathrm{occurrence})$, an EMRI occurs during the LISA observing window. If an EMRI occurs, we sample the remaining parameters from Table~\ref{tab:Param_dist} and use them as input for the Numerical Kludge code, which outputs $\{\mathrm{time}, h_+, h_\times \}$.
 
\subsection{\label{sec:expeep}Peep Signals}

An example peep waveform can be seen in Figure~\ref{fig:test_wave_char} (top). The parameters for this peep waveform are: $a=0.815535M$, $p_0=113.194144M$, $e_0=0.992424$, $\iota_0=2.363436^\circ$, $log_{10}\ M=5.2\ M_\odot$, $\mu=18.152765\ M_\odot$, $\theta_s=0.581082^\circ$, $\phi_s=14.558132^\circ$, $z=0.6$, and $dt=15 \mathrm{s}$. The waveform was modeled with 4 years of an EMRI orbit in the source frame, which is brought into the detector frame by multiplying the time by $(1 + z)$ and dividing the strain amplitude by $(1 + z)$. Since $z=\Delta \lambda/\lambda$, the natural ratio for redshifting quantities is $(1+z)$ which is simply $\lambda + \Delta \lambda/\lambda$ or the ratio of the redshifted wavelength to the emitted wavelength. Waveforms are stretched by this factor as they travel through the cosmos from a time at redshift z to the present and, in addition, such sources are dimmed (hence the division of the strain amplitude) by an amount that depends on the same factor because the density of gravitons per unit volume decreases as a result of the expansion of space \cite{hogg2000distance}. Redshifting our signal results in 6.4 years of data. Here we present the first 4 years of the signal in the detector frame. During this time, the EMRI emits sixteen peeps of gravitational radiation. The amplitude shift in the waveform is due to the inclination of the orbit with respect to the spin axis of the MBH. A 2D representation of the first four months of this orbit can be seen in Cartesian coordinates in Figure~\ref{fig:peep_cartesian}.

Once we have the gravitational waveform, we can pass it through \texttt{fastlisaresponse} with a randomly drawn sky location from Table~\ref{tab:Param_dist}. The result is Figure~\ref{fig:test_wave_char} (middle), where we show the A and E channel outputs from the response function. We can then take a Fast Fourier Transform (FFT) to obtain $\tilde{A}(f)$ and $\tilde{E}(f)$. Next, because the peep is not entirely within the LISA bandwidth, we filter out the frequencies that are lower than LISA's frequency range. We can then compute the characteristic strain by plugging our values into Equation~\ref{eqn:char_strain} (where $\tilde{X}$ denotes the $\tilde{A}$ or $\tilde{E}$ channels, and $f$ is the frequency array corresponding to $\tilde{X}$ with a lower bound of $10^{-5}$ Hz) and then plot it against the LISA noise curve for the given channel which gives us Figure~\ref{fig:test_wave_char} (bottom). Finally, we can compute the SNR of the signal of our $\tilde{A}$ and $\tilde{E}$ channels by using Equation~\ref{eqn:SNR}\cite{babak2021lisasensitivitysnrcalculations} (where $S_n(f)$ corresponds to the LISA sensitivity curve for 2nd generation TDI, generated using \texttt{LISAAnalysisTools} \cite{michael_katz_2024_10930980}, assuming stationary, Gaussian noise and ESA trailing orbits, evaluated over the frequency array corresponding to $\tilde{X}$, and where $f_{\mathrm{max}}$ is the upper bound of that frequency array). For this example, it is well below the detectability threshold with SNR for the A and E channels being $\sim 0.0012$ and the combined SNR is $\sim 0.0017$.

\begin{equation}
\label{eqn:char_strain}
h_c(f) = 2 f\left(\left|\tilde{X}(f)\right|^2\right)^{1 / 2}
\end{equation}

\begin{equation}\label{eqn:SNR}
    \mathrm{SNR}^2 = 4\mathrm{Re}\left(\int_0^{f_{\mathrm{max}}}df\frac{\tilde{X}\tilde{X}^*}{S_n(f)}\right)
\end{equation}

\subsection{Peep Background}

Following the methods described in \S\ref{sec:results}, we completed the steps to obtain a total of 1470 unique time domain A and E channel waveforms. The number of unique EMRI signals per year is comparable to model M4 from Babak et al. (2017) \cite{babak_science_2017}, which had a similar plunge rate and CO mass. Being comparable to the models in Babak et al. (2017) makes our catalog complementary to existing studies, while offering a broader perspective due to our varied physical assumptions on EMRI parameters.

For \textbf{Assumption 1} and \textbf{Assumption 2} backgrounds, we performed new draws of our parameter distributions and recomputed the waveforms, so they are unique between the two backgrounds. As the data is computed in the source frame and then redshifted into the detector frame, the individual waveforms become irregularly sampled in time. To enable consistent stacking and analysis of our combined samples, we interpolate the redshifted data onto a uniform 15-second cadence in the detector frame using \texttt{scipy.interpolate.UnivariateSpline} which is a 1-dimensional spline fit that preserves the physical features of the waveform while minimizing numerical artifacts. We then iteratively add together the individual channels (A \& E) to generate a gravitational wave background. We can then use Equation~\ref{eqn:char_strain} to convert these backgrounds into the characteristic strain, and then plot the background over the \texttt{A1TDISens} or \texttt{E1TDISens} sensitivity curves. Finally, we can calculate the SNR of our new background by utilizing Equation~\ref{eqn:SNR}.

\begin{figure*}[t]
  \centering
  \includegraphics[width=\textwidth]{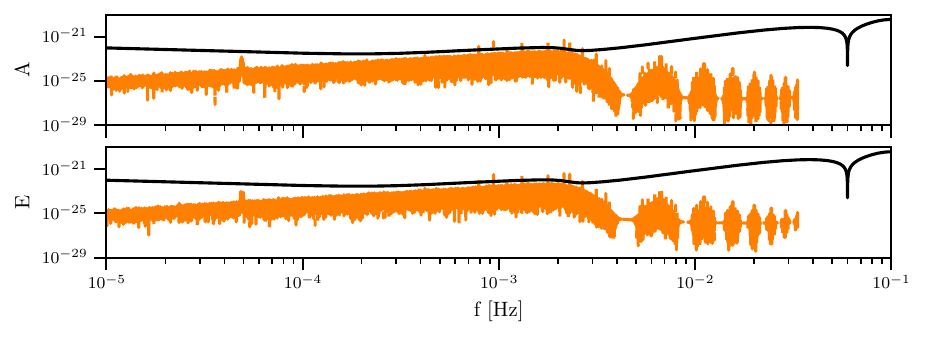}
    \caption{\textbf{Assumption 1 Background}: 1,470 total peeps that were passed through the LISA response function from \texttt{fastlisaresponse}\cite{lisaongpu1,lisaongpu2} and then iteratively combined into a single background for each out the LISA TDI channels, A and E. We then took a FFT of each of the channels and plugged the values into Equation~\ref{eqn:char_strain} to obtain the characteristic strain of the signal. The LISA sensitivity curve for each channel, A1TDISens and E1TDISens, (black) are then plotted over the background (orange). The amplitude of the background is just below the sensitivity curve and the total SNR in each channel is of order  $\sim 0.2$ for the A and E channels, and a combined SNR of $0.33$.}
    \label{fig:aetwave4-1}
\end{figure*}

The gravitational wave background for \textbf{Assumption 1} is shown in Figure~\ref{fig:aetwave4-1}. The respective signals in each channel are just below the LISA instrumental noise and have SNRs  of order $\sim 0.2$ for the A and E channels, and a combined SNR of $0.33$. 

\begin{figure*}[t]
    \centering
    \includegraphics[width=\textwidth]{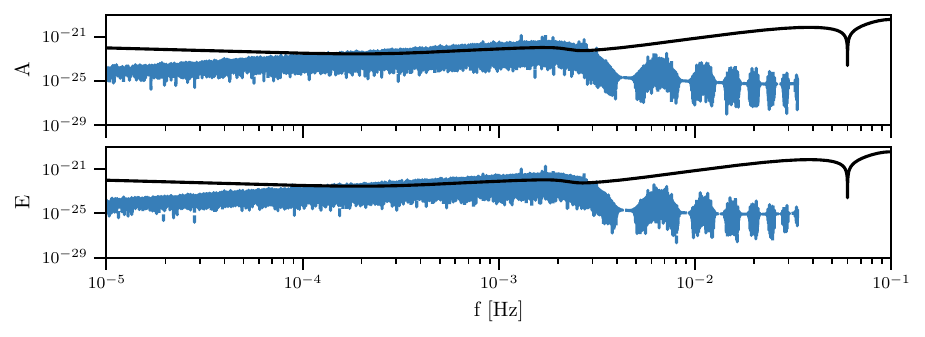}
    \caption{\textbf{Assumption 2 Background}: 1,470 resampled peeps with lower bounds on $p_0$ and $e_0$ that were passed through the LISA response function from \texttt{fastlisaresponse}\cite{lisaongpu1,lisaongpu2} and then iteratively combined into a single background for each out the LISA TDI channels, A and E. We then took a FFT of each of the channels and plugged the values into Equation~\ref{eqn:char_strain} to obtain the characteristic strain of the signal. The LISA sensitivity curve for each channel, A1TDISens and E1TDISens, (black) are then plotted over the background (blue). The signal, while still well below the LISA sensitivity curve is an order of magnitude larger than that of Assumption 1. The total SNR in each channel is of order $\sim 1.7$, and a combined SNR of $2.4$.}
    \label{fig:aetwave4-2}
\end{figure*}

We can then perform the same steps to the above and determine the gravitational wave background for \textbf{Assumption 2}, which has different lower bounds on the semi-latus rectum and orbital eccentricity ($8M \leq p_0 \leq 120M$ and  $0.9 \leq e_0 \leq 0.999999$). The characteristic strains are plotted against the LISA sensitivity curve for the individual (A \& E) channels and are shown in Figure~\ref{fig:aetwave4-2}. The signals are just above the LISA sensitivity curve, resulting in a background SNR that is an order of magnitude larger than assumption 1. The A and E channels are of order  $\sim 1.7$, and a combined SNR of $2.4$.

\begin{figure*}[t]
    \centering
    \includegraphics[width=\textwidth]{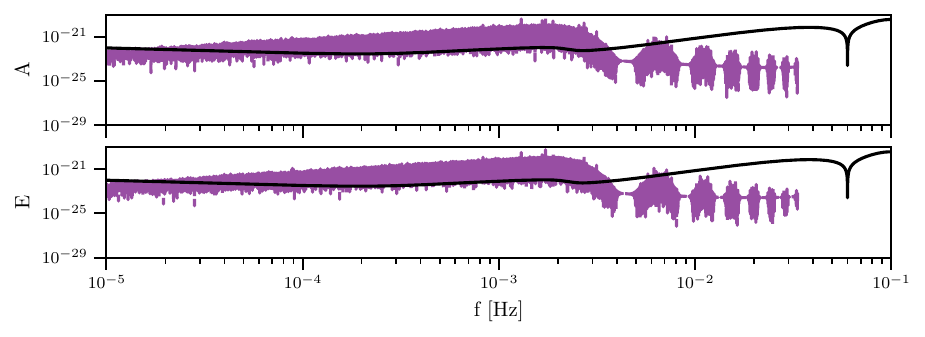}
    \caption{\textbf{Assumption 3 Background}: 1,470 resampled peeps with lowered bounds on $p_0$ and $e_0$ with the added assumption that each peep in the sample occurred 1000 times at a given sky location, effectively making the background made up of $1.47\times10^6$ peeps. We computed this by taking background 2 and multiplying by $\sqrt{1000}$, which is a good estimate for the incoherent sum. This assumption comes from Amaro-Seoane et al. (2024), which estimates that there could be thousands of these signals in a Milky Way-type galaxy. Each of the waveforms was then passed through the LISA response function from \texttt{fastlisaresponse}\cite{lisaongpu1,lisaongpu2} and then iteratively combined into a single background for each of the LISA TDI channels, A and E. We then took an FFT of each of the channels and plugged the values into Equation~\ref{eqn:char_strain} to obtain the characteristic strain of the signal. The LISA sensitivity curve for each channel, A1TDISens and E1TDISens (black), are then plotted over the background (purple). The final output differs from the assumptions put forth by \cite{seoane2024mono}, which estimated that the incoherent sum of highly eccentric early-stage EMRIs would reach characteristic strain values as high as $10^{-16}$ and frequencies as high as $3\times 10^{-2}$ Hz. Whereas our output reaches an amplitude of $10^{-20}$ at its largest. The final characteristic strain in the A and E channels is now larger than the sensitivity curve across frequency and has significantly higher SNRs. The total SNR in each channel is of order $\sim 55$ for the A and E channels and a combined SNR $\sim77$.}
    \label{fig:aetwave4-3}
\end{figure*}

For the third background, we followed more abundant estimates of highly eccentric EMRIs early in their inspiral from \cite{seoane2024mono}, we take the results from background 2, and we multiplied the results by $\sqrt{1000}$, effectively accounting for 1000 incoherent signals at a given sky location. This was done as it would be too computationally expensive to compute fully. The number of sources that are implied from this estimation would be comparable to that of M7 or M12 from Babak et al. (2017) \cite{babak_science_2017}. The results of this can be seen in Figure~\ref{fig:aetwave4-3}. This is the background that provides the most interesting result, as the SNR is much larger than the LISA sensitivity curve. The A and E channels have an SNR of order $\sim 55$ each and a combined SNR of $\sim77$.

\begin{figure*}[t]
    \centering
    \includegraphics[width=\textwidth]{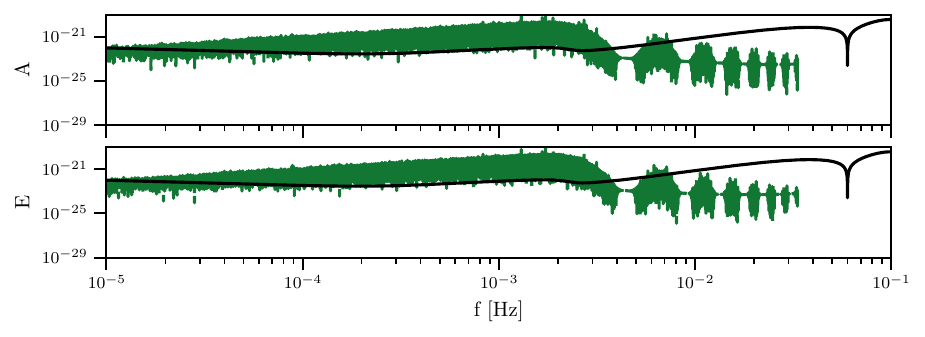}
    \caption{\textbf{Assumption 4 Background}: 1,470 resampled peeps with lower bounds on $p_0$ and $e_0$, while estimating the contribution of signals that have remained individually unresolvable over the past $10^5$ years. This effectively represents a background equivalent to $\sim5.21\times10^6$ sources evolving over $10^5$ years. The estimate is obtained by scaling the background from \textbf{Assumption 2} by $\sqrt{3545}$, which approximates the incoherent sum of additional bursts in each peep. The factor of 3545 was derived by computing orbital periods for all sources, excluding those with periods $\leq 4$ years, and passing the values through Equation~\ref{eqn:factor}, where $\mathcal{F}$ is the factor, $\mathcal{T}_{\mathrm{ref}}$ is the reference time (the maximum orbital period from our parameter distribution), and $\mathcal{P}_{\mathrm{orb}}$ is the average orbital period of our EMRIs with periods greater than 4-years. Each waveform in Background 2 was processed through the LISA response function using \texttt{fastlisaresponse}\cite{lisaongpu1,lisaongpu2}, and then iteratively combined into a single background for each of the LISA TDI channels, A and E. We then took an FFT of each of the channels and plugged the values into Equation~\ref{eqn:char_strain} to obtain the characteristic strain of the signal. The LISA sensitivity curve for each channel, A1TDISens and E1TDISens (black), are then plotted over the background (green). This is our largest background made of individiaully unresolvable sources, exceeding the sensitivity curve across nearly the entire frequency range and producing significantly higher SNRs than Background 3, with SNR $\sim 100$ in the A and E channels and a combined SNR of $\sim 145$.}
    \label{fig:aetwave4-4}
\end{figure*}

\begin{figure*}[t]
    \centering
    \includegraphics[width=\textwidth]{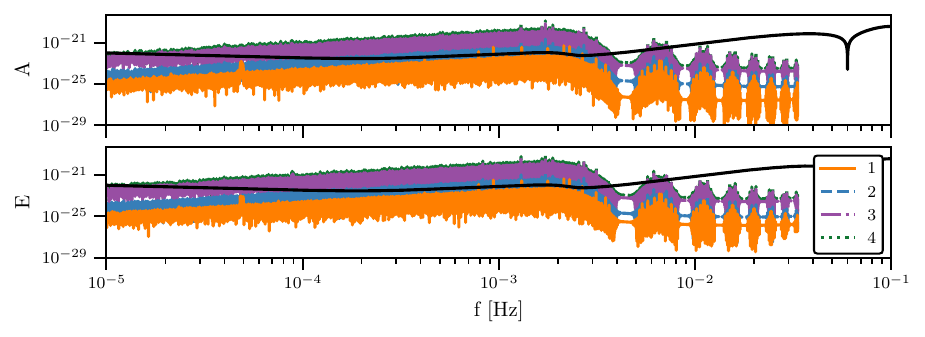}
    \caption{Comparison between the backgrounds generated from each of the three assumptions described in Section~\ref{sec:results} and shown in Figures~\ref{fig:aetwave4-1}-\ref{fig:aetwave4-4}. \textbf{Assumption 1} is shown in orange, \textbf{Assumption 2} is shown in blue, \textbf{Assumption 3} is shown in purple, and \textbf{Assumption 4} is shown in green. Backgrounds 1 and 2 are right at the level of the sensitivity curve and are not likely to obscure otherwise detectable sources. However, the more abundant cases, Backgrounds 3 and 4, are well above the sensitivity curve.}
    \label{fig:fulbackcomp}
\end{figure*}

To construct the final background, we account for the contribution from highly eccentric EMRIs that formed prior to the 4-year window but have orbital periods so large that they would otherwise be excluded. Directly modeling these systems over $\sim10^5$ years for all 1470 sources is computationally prohibitive, so we must approximate. We take the maximum orbital period possible from our parameter distribution as the reference timescale,  $\mathcal{T_{\mathrm{ref}}}=10^5$ years. Then we aim to obtain an estimated factor ($\mathcal{F}$) for our average number of additional orbital periods within our reference timescale. First, we excluded signals with orbital periods ($\mathcal{P_{\mathrm{orb}}}$) shorter than 4 years, as these are likely to evolve into individually resolvable sources within $10^5$ years. We then computed the multiplicative factor using Equation~\ref{eqn:factor} using our remaining systems and obtained a value of 3545 additional orbits per source using the parameters from \textbf{Assumption 2}. To account for partial constructive and destructive interference among signals, we scale the background from \textbf{Assumption 2} by $\sqrt{3545}$. The resulting spectrum, shown in Figure~\ref{fig:aetwave4-4}, produces an even stronger gravitational wave background than \textbf{Assumption 3}, with SNRs of order $\sim 100$ for the A and E channels and a combined SNR of $\sim 145$. A comparison of the characteristic strain for all four backgrounds is presented in Figure~\ref{fig:fulbackcomp}.

\section{\label{sec:conc}Conclusion\protect}

Extreme Mass Ratio Inspirals are a very important source for LISA due to the many tests of general relativity that can be performed, as well as potentially providing useful data such as accurate mass and spin measurements of both stellar mass and massive black holes. The later stages of EMRIs are much more likely to be detectable. While several studies have been performed on the potential signal confusion produced by ``undetectable" types of EMRIs, as far as we have been able to find, very few have explored the very highly eccentric regime examined in this study, and none with realistic population estimates from Illustris, and updated estimates of capture parameters and capture rates. 

Utilizing the numerical kludge allows us to capture many of the important relativistic effects of these highly eccentric EMRIs while maintaining a reasonable amount of computation speed. Many current waveform models lose a tremendous amount of accuracy as the CO becomes very close to the MBH (smaller semi-latus rectum value), which is why our study focused on the EMRIs that were farther out and thus less likely to be detectable. Since these orbits also do not evolve appreciably in a 4-year window, we also modeled the orbits without this orbital evolution.

In this study, we used Illustris, a cosmological simulation, to determine the SMBH mass function. We then combined this information with EMRI formation rates based on mass ratio and estimates of EMRI capture parameters to obtain the number of captures as a function of redshift and mass for these highly eccentric ``peeping" EMRIs. With this study, we observed 1470 gravitational wave peeps across a large parameter space out to redshifts of up to $z = 3$ for four years in the source frame. As the amplitude of the waveforms is inversely proportional to distance, for sources beyond $z = 3$, we believe that there would be a minimal contribution to SNR and therefore can be reasonably excluded from our study. Our number of unique EMRIs is comparable to model M4 from Babak et al. (2017) \cite{babak_science_2017}. These signals were then redshifted and combined for four years of data in the detector frame and then passed through \texttt{fastlisaresponse} for the three different background assumptions shown in Figure~\ref{fig:fulbackcomp}. 

The first two assumptions resulted in combined backgrounds that just reach the sensitivity curve for LISA. In contrast, the third background, based on the assumptions from \cite{seoane2024mono}, posits that for each forming peep, we have 1000 additional peeps, which resulted in a total combined SNR of $\sim 77$. The fourth background then provides an estimate based on our realistic population rates, of an upper limit on this background based on these signals recurring over the course of $10^5$ years resulting in a total combined SNR of $\sim 145$. Since the first two backgrounds are based on our population rate estimates, we consider them more realistic. While they are unlikely to obscure otherwise detectable sources, the third and fourth backgrounds, due to their significantly more abundant populations, produces signals strong enough to potentially mask sources. Backgrounds 3 and 4 are estimations and should be taken as such. A thorough examination involving those signals which evolve to be individually detectable, or those which re-scatter while approaching apoapsis is beyond the scope of this work and will be revisited in a future analysis.

It seems likely that these contrasting background assumptions represent lower and upper bounds for the number of highly eccentric early-stage ``peep"-like EMRIs that will be observed by LISA. The true figure may be greater than one per galaxy, but it is likely considerably less than 1,000 per galaxy, and additionally sources which have formed prior to LISA's 4-year observing window must also be considered. Since our results indicate that any increase upon one per galaxy is likely to impinge upon the LISA sensitivity curve in such a way as to compromise successful extraction of detectable sources, it would seem that ``peep"-like orbits warrant further investigation. In particular, more closely modeling the assumptions of background assumption 4, would provide valuable insights into the entire EMRI spectrum and not just the highly eccentric undetectable sources being examined in this work.

Detectability is not a strict requirement for astrophysical signals to cause confusion noise for other signals, as is the case for the millions of individually unresolvable galactic binaries that cause a foreground resulting in a modification to the noise curve \cite{barack_confusion_2004, bonetti_gravitational_2020}.
Instead, one should compare the number of frequency bins to the number of signals and their behavior in the time-frequency domain.
In this case, we only have $\mathcal{O}(10^3)$ signals as compared to the number of bins $\Omega / \Delta f \sim 0.5\times10^6$, where $\Omega\sim 4\times10^{-3}$~Hz is the bandwidth affected by these backgrounds, and $\Delta f = 1/T\sim 8\times10^{-9}$~Hz is the frequency bin width with 4 years of observation.
Only in the cases of \textbf{Assumptions 3 and 4} do we find a number of signals that is similar to the number of bins with this observation time span.
However, if many of the harmonics contribute significantly for these signals, there could be many time-frequency crossings per signal, potentially resulting in signal confusion noise, and this could be investigated in a follow-up study using realistic, highly eccentric EMRI waveforms alongside other signal types.

Some potential ways to improve on this work will be to supplement our data with a more accurate code for sources that begin at $p=15M$ and inward, which have larger amplitudes and thus contribute to the background more. Another regime to look at is the potential of peeps in the galactic center, which may contribute to a galactic foreground of potentially unresolvable sources, though because they are closer, they too will have much larger amplitudes. Lastly, with LISA's recent adoption, there is likely to be a new constellation dynamics model in the coming years, which may impact the results of the waveform in the individual A and E channels.

This paper utilized \texttt{SciPy} \cite{scipy}, \texttt{NumPy} \cite{numpy}, \texttt{AstroPy} \cite{astropy:2013,astropy:2018,astropy:2022}, \texttt{Matplotlib} \cite{matplotlib}, \texttt{Mathematica} \cite{Mathematica}, and the \texttt{Black Hole Perturbation Toolkit} \cite{BHPToolkit}.

\section{Acknowledgements\protect}

D. J. O.’s work is supported by NSF Physics Frontiers
Center Award No. 2020265. A. D. J. acknowledges support
from the Caltech and Jet Propulsion Laboratory President’s
and Director’s Fund. J. B. and L. J.’s contributions to this
work were supported by the National Science Foundation
under Award No. OIA-1557417. K. G. acknowledges
support from research Grant No. PID2020–1149 GB-I00
of the Spanish Ministerio de Ciencia e Innovación. This
research is supported by the Arkansas High Performance
Computing Center which is funded through multiple
National Science Foundation grants and the Arkansas
Economic Development Commission.

\section{Data Availability}

The data for this work is openly available on Zenodo \cite{oliver_2026}.

\bibliographystyle{apsrev4-2}
\bibliography{peep2}

\end{document}